\documentclass{emulateapj}
\newcommand{\bl}[1]{\mbox{\boldmath$ #1 $}}
\newcommand{\be}{\begin{equation}}
\newcommand{\ee}{\end{equation}}
\newcommand{\cs}{c_{\rm s}}
\newcommand{\csq}{c_{\rm s}^2}

\newcommand{\msun}{M_{\sun}}

\newcommand{\mdot}{\dot{M}}
\newcommand{\mdotavg}{\langle \dot{M} \rangle}

\newcommand{\mdiskavg}{\langle M_{\rm d} \rangle}
\newcommand{\mstaravg}{\langle M_* \rangle}

\slugcomment{Accepted by Astrophysical Journal Letters}
\shorttitle{T Tauri Disk Accretion}
\shortauthors{E. I. Vorobyov and Shantanu Basu}
\begin{document}

\title{Mass accretion rates in self-regulated disks of T Tauri stars}

\author{E. I. Vorobyov\altaffilmark{1,}\altaffilmark{2},
Shantanu Basu\altaffilmark{3}}
\altaffiltext{1}{Institute for Computational Astrophysics, Saint Mary's University,
Halifax, B3H 3C3, Canada; vorobyov@ap.smu.ca.} 
\altaffiltext{2}{Institute of Physics, South Federal University, Stachki 194, Rostov-on-Don, 
344090, Russia.}
\altaffiltext{3}{Department of Physics and Astronomy, University of Western Ontario,
London, Ontario,  N6A 3K7, Canada; basu@astro.uwo.ca.}

\begin{abstract}{}
We have studied numerically the evolution of protostellar disks around 
intermediate and upper mass T Tauri stars ($0.25~M_\odot < M_\ast< 3.0~M_\odot$) that have formed 
self-consistently from the collapse 
of molecular cloud cores. In the T Tauri phase, 
disks settle into a self-regulated state, 
with low-amplitude nonaxisymmetric 
density perturbations persisting for at least several million years.
Our main finding is that the global effect of gravitational torques due to these perturbations 
is to produce disk accretion rates that are of the correct magnitude to explain observed accretion
onto T Tauri stars. 
Our models yield a correlation between accretion rate $\mdot$ and stellar mass $M_\ast$
that has a best fit $\mdot \propto M_\ast^{1.7}$, in good agreement with recent observations.
We also predict a near-linear correlation between the disk accretion rate
and the disk mass. 
\end{abstract}


\section{Introduction}

Gaseous circumstellar disks are known to extend anywhere from several to 
perhaps hundreds of AU around T Tauri stars (TTS), due to the 
detection of millimeter and submillimeter emission from the associated
dust \citep[e.g.,][]{Beckwith,Andrews}. It has 
recently been recognized that such disks also exist
around brown dwarfs (BD), 
with masses as small as $0.01~\msun$, but with comparable 
disk-to-star mass ratios as found for TTS \citep{Klein,Scholz1}.
Spectroscopic observations demonstrate that BD share similar 
accretion properties with TTS. 
In particular, both have a large scatter in the mass accretion rate
($2-3$ orders of magnitude) at a given stellar mass 
\citep{Scholz2} and, when data for BD and TTS are taken {\it together}, 
the accretion rates $\dot{M}$ show a strong direct dependence on 
stellar masses $M_{\ast}$, with an approximate scaling
$\dot{M}\propto M_{\ast}^2$ \citep[e.g.,][]{Muzerolle,Muzerolle2,Mohanty,Natta}.

The origin of the above relation is uncertain. 
\citet{Padoan}
have argued that this relation is a consequence of Bondi-Hoyle accretion 
from the large-scale gas distribution in the parent cloud. 
Such a scenario neglects the importance of disk physics in determining
the accretion rate,
and also fails to explain the similarity of accretion
rates onto TTS located within molecular clouds and H~II regions 
\citep{Hartmann}. An alternative idea is that the accretion onto the stellar surface 
is controlled by viscous disk evolution. 
\citet{AA} and \citet{Hartmann} have invoked different initial input
parameters or scaling relations into the standard viscous disk 
evolution models of \citet{Hartmann98} in order to explain the 
approximate $\dot{M}\propto M_{\ast}^2$ scaling.
These models treat disks as isolated smooth axisymmetric structures
that evolve due to an unspecified source of turbulent viscosity
that has ad-hoc spatial dependence as well as temporal independence.
\citet{Dullemond} have taken a more fundamental approach of linking the
disk evolution to the properties of the collapsing core from which it forms,
but still relies upon the ad-hoc $\alpha$-viscosity prescription to model the 
angular momentum transport within the disk. 

In this Letter, we take the basic approach of studying the mass accretion
rate within disks that have formed self-consistently due to the collapse
of cloud cores.
The disks are actually profoundly nonaxisymmetric when formed
in this manner, and
experience significant envelope-induced gravitational instability 
and accretion bursts during their {\it early} evolution 
\citep[][hereafter VB05,VB06, respectively]{VB1,VB2}.  
Here, we explore the {\it late} evolution of such self-consistently
formed disks, when their evolution is governed by low amplitude
nonaxisymmetric density perturbations
\citep[][hereafter VB07]{VB3}. The gravitational torques produced 
by these perturbations are sufficient to explain the observed magnitudes
and scatter of accretion rates of TTS, and even approximately fit the observed 
$\dot{M}-M_{\ast}$ relation.

\section{Model description}

We use the thin-disk approximation to compute the evolution of rotating, 
gravitationally bound cloud cores. This allows efficient calculation of 
the long-term evolution of a large number of models.
The derivation of the relevant equations, 
details of the numerical code and tests are given in VB06.
The basic equations of mass and momentum transport are

\begin{eqnarray}
\label{cont}
 \frac{{\partial \Sigma }}{{\partial t}} & = & - \nabla _p  \cdot \left( \Sigma \bl{v}_p 
\right), \\ 
\label{mom}
 \Sigma \frac{d \bl{v}_p }{d t}  & = &  - \nabla _p {\cal P}  + \Sigma \bl{g}_p \, ,
\end{eqnarray}
where $\Sigma$ is the mass surface density, 
${\cal P}$ is the vertically integrated gas pressure,
$\bl{v}_p$ is the velocity in the
disk plane, $\bl{g}_p$ is the gravitational acceleration in the disk plane,
and $\nabla_p$ is the gradient along the planar coordinates of the disk.
Equations~(\ref{cont}) and (\ref{mom}) are closed with a barotropic equation
that makes a transition from isothermal to adiabatic evolution at 
$\Sigma = \Sigma_{\rm cr} = 36.2$ g cm$^{-2}$.
This approach bypasses the detailed cooling and heating mechanisms,
but provides a good fit to the density-temperature relation in collapsing 
cloud cores (see VB06). In the late phase of disk evolution, the
stellar irradiation is expected to be a significant source of heating
and is neglected in our simulation. Much of the region of enhanced
temperature would fall within our ``sink cell'' (see below), but our
calculated temperatures at $\sim 10$ AU are also somewhat lower than 
found in models of stellar irradiation onto flared passive disks
\citep{Chiang}. Further details of such a comparison are given by
VB07.  

Equations~(\ref{cont}) and (\ref{mom}) are solved in polar coordinates
$(r, \phi)$ on a numerical grid with
$128 \times 128$ points. The radial points are logarithmically spaced.
The innermost grid point is located at $r=5$~AU, and the size of the 
first adjacent cell is 0.3~AU.  We introduce a ``sink cell'' at $r<5$~AU, 
which represents the central protostar plus some circumstellar disk material, 
and impose a free inflow inner boundary condition.  
The outer boundary is such that the cloud has a constant mass and volume.

The gas has a mean molecular mass $2.33 \, m_{\rm H}$ and is initially
isothermal with temperature $T$ ($=10$ K) and isothermal sound speed 
$\cs$. 
The initial distributions of $\Sigma$ and angular velocity $\Omega$ 
are those characteristic of a collapsing axisymmetric magnetically
supercritical core \citep{Basu}:
\begin{equation}
\Sigma={r_0 \Sigma_0 \over \sqrt{r^2+r_0^2}}\:,
\label{dens}
\end{equation}
\begin{equation}
\Omega=2\Omega_0 \left( {r_0\over r}\right)^2 \left[\sqrt{1+\left({r\over r_0}\right)^2
} -1\right].
\end{equation}
The asymptotic $r^{-1}$ power-law dependence of these quantities is a 
robust result of rotating cloud collapse simulations
\citep{Norman,Narita}. These profiles have the property 
that the specific angular momentum $j=\Omega r^2$ is a linear 
function of the enclosed mass $m$.
The scale length $r_0 = k \csq /(G\Sigma_0)$, where $k\simeq 1$
\citep{Basu}. For the models in this paper, we adopt $k = \sqrt{2}/\pi$.
These initial profiles are characterized by the important
dimensionless free parameter $\gamma \equiv  \Omega_0^2r_0^2/\csq$. 
The asymptotic ($r \gg r_0$) ratio of centrifugal to gravitational
acceleration has magnitude $\sqrt{2}\,\gamma$ \citep[see][]{Basu} and 
the centrifugal radius of a
mass shell initially located at radius $r$ is estimated to be
$r_{\rm cf} = j^2/(Gm) = \sqrt{2}\, \gamma r$. 
Since the enclosed mass $m$ is a linear function of $r$ at large radii,
this also means that $r_{\rm cf} \propto m$.

We present results from three sets of models in this Letter, each
with a different value of $\gamma$.  The standard model has 
$\gamma = \gamma_1 = 1.2 \times 10^{-3}$ based on typical values
$\cs = 0.19$ km s$^{-1}$, $\Sigma_0 = 0.12$~g~cm$^{-2}$, and
$\Omega_0 = 1.0$~km~s$^{-1}$~pc$^{-1}$. The outer radius is taken to
be $r_{\rm out} = 0.04$ pc, and the total cloud mass is $0.8\,\msun$. 
Other models with $\gamma = \gamma_1$ but different mass are generated
by varying $r_0$ and $\Omega_0$ so that their product is constant. All
clouds are characterized by the same ratio $r_{\rm out}/r_0\approx 6.0$. 
To generate the second set of models, with $\gamma = \gamma_2 = 2.3 \times 10^{-3}$, we set
$\Omega_0 = 1.4$~km~s$^{-1}$~pc$^{-1}$ and all other quantities the
same as in the standard model with $\gamma= \gamma_1$. Models of
varying mass are then generated in the same manner as for the
$\gamma_1$ models. The third set of models, with $\gamma=\gamma_3
= 3.4 \times 10^{-3}$,
are also obtained in this way,
by first using $\Omega_0 = 1.7$~km~s$^{-1}$~pc$^{-1}$.
Overall, there are 6 models with $\gamma = \gamma_1$, 14 models with
$\gamma = \gamma_2 \simeq 2\, \gamma_1$, 
and 12 with $\gamma = \gamma_3 \simeq 3\, \gamma_1$.
The range of initial cloud
masses amongst our models is $0.25\,\msun-3.0\,\msun$.

The numerical simulations start in the prestellar phase and continue into 
the late accretion phase, long after the formation of a protostar and a protostellar disk.
The disk evolution is followed for approximately 3~Myr after the formation of the protostar. 
In the early phase, when the infall of matter from the surrounding envelope is substantial, mass is
transported inward by the gravitational torques from spiral arms that are a manifestation of the
envelope-induced gravitational instability in the disk 
(VB05; VB06).
In the late phase, when the gas reservoir
of the envelope is depleted, the distinct spiral structure is replaced by ongoing irregular 
nonaxisymmetric density perturbations in the disk. 
These perturbations are confined to the disk, of size $\sim 100$ AU, and
are not affected by the computational boundary at $\sim 5,000-10,000$ AU. 
Instead, their longevity is enhanced by swing amplification at
the disk's sharp {\it physical} boundary (see VB07).
We find that the net effect from these density perturbations is a 
residual non-zero gravitational torque.
In particular, the net 
gravitational torque in the inner disk tends to be negative during first several million 
years of the evolution, while the outer disk has a net positive gravitational torque.
The inward radial transport of matter due to the negative torque, 
which is produced {\it self-consistently} in our numerical hydrodynamic modeling, 
is the essence of the accretion mechanism in our model.

Our model accretion rates are consistent with 
typical accretion rates for intermediate and upper-mass TTS 
($0.25~M_\odot < M_\ast< 3.0~M_\odot$). We do not consider objects with masses
below $0.25~M_\odot$, because the numerical noise generated by the inner boundary becomes 
comparable to the amplitude of density perturbations in compact disks around extremely 
low-mass objects ($<0.1~M_\odot$). 

\begin{figure*}
 \centering
  \includegraphics{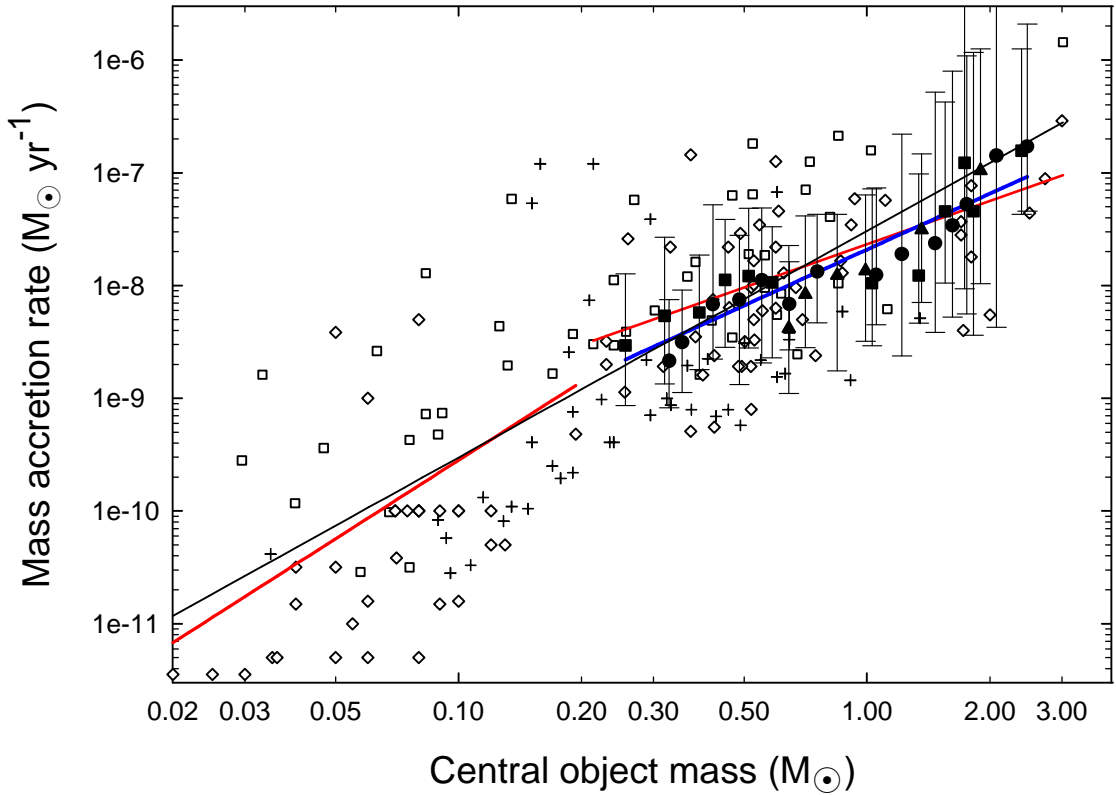}
      \caption{Mass accretion rates versus stellar masses. The open
      diamonds represent measurements of TTS and BD from
\citet[][ and references therein]{Muzerolle2}; the open squares and 
crosses represent the confirmed detections and upper limits, 
respectively, compiled by \citet{Natta}. 
      The filled triangles, filled circles, and filled squares show the time-averaged 
      quantities for the models with $\gamma = \gamma_1, \gamma_2,\: {\rm and} \: \gamma_3$, 
      respectively. 
      The bars represent mean positive/negative deviations from the time-averaged accretion rates
      in each model. The blue line ($\propto M_\ast^{1.7}$) is the least-squares fit to the
      model data. The black line ($\propto M_\ast^{2.0}$) is the least-squares fit to the
      observed confirmed detections, both TTS and BD. The left and right red lines are 
      the least-squares fits to objects with masses in the range $(0.02-0.2)~M_\odot$ and 
      $(0.2-3.0)~M_\odot$, respectively.}
         \label{fig1}
\end{figure*}

\section{Accretion rates}
Figure~\ref{fig1} shows confirmed
detections for the mass accretion rate $\dot{M}$ and stellar mass $M_\ast$ 
for TTS and BD of age $0.5-3$~Myr from two recent observational
compilations. The open diamonds represent measurements, mostly in Taurus, that
have been compiled by \citet[][ and references therein]{Muzerolle2}; the open
squares represent detections in $\rho$ Oph obtained by \citet{Natta} and crosses 
represent their upper limits to nondetections. 
We have excluded objects in the compilation of \citet{Muzerolle2}
that were later observed by \citet{Natta}. The least-squares best fit to the observational 
data, both TTS and BD (upper limits excluded), has an exponent $2.0 \pm 0.1$ 
and is shown in Figure~\ref{fig1} by a black line. This value is often quoted in the literature (see \S\ 1). However, we believe that taking a 
best fit over the whole mass range of BD and TTS may be misleading.
Indeed, if we consider separately the intermediate and upper-mass TTS
($0.25~M_\odot < M_\ast< 3.0~M_\odot$) and lower-mass TTS plus BD ($0.02~M_\odot<M_\ast<0.25~M_\odot$),
then the least-squares best fits to each data sample are distinct. 
In particular,
for lower-mass TTS and BD we obtain an exponent $2.3 \pm 0.6$ (left red line), whereas for 
the intermediate and upper-mass TTS we find a much smaller exponent $1.3 \pm 0.3$ (right red line).

The above data hints that different mechanisms may be responsible
for accretion as one moves along the sequence of stellar masses. 
However, the observational method of determining $\dot{M}$ also typically
varies across the mass sequence, with significant but differing uncertainties.
For TTS, the primary determinant of $\dot{M}$ at the stellar surface is
UV excess and veiling \citep[see, e.g.][]{Muzerolle}, while for BD the
primary means is the fitting of emission line profiles.
\citet{Muzerolle} point out that the latter is considered less accurate
than UV excess measurements; it suffers from uncertainties regarding
optical depth, rotation, and inclination effects \citep[see also discussion in][]{Mohanty}. 
Quantitative estimates of error bars for either technique 
are not available in the literature.
However, the agreement between the two methods is within a factor
$\sim 3-5$ \citep{Muzerolle} where comparable, and this is much smaller
than the spread of observed $\dot{M}$ for a given central object mass, 
implying that the spread of values is primarily due to physical 
effects of disk age and initial conditions.
We also note that the data set of \citet{Natta} is based entirely on 
emission line estimates of accretion onto BD and TTS.

Figure \ref{fig1} also shows the 
time-averaged mass accretion rates $\mdotavg$ and time-averaged stellar masses $\mstaravg$ 
for our models with $\gamma = \gamma_1, \gamma_2,\: {\rm and} \: \gamma_3$, respectively.
The mass accretion rate $\dot{M}(t)= - 2 \pi r v_r \Sigma$,
where $v_r$ is the inflow velocity of gas through the sink cell and $r=5$~AU.
For most models, the time average is taken between 0.5~Myr and 3.0~Myr after the formation
of the protostar. However, protostars with masses above $2.0~M_\odot$ may enter the T Tauri 
phase (class~II) when they are older than 0.5~Myr. To exclude a possible input 
from class~0/class~I sources in such cases, we start the time average only when the envelope mass 
has dropped below $10\%$ of the initial cloud mass. The best fit to our model data points is 
\begin{equation}
\mdotavg  = 10^{-7.7} \; \mstaravg^{1.7}.
\label{relation4}
\end{equation} 
and is represented in Figure~\ref{fig1} with a blue line. The
least-squares method generates an uncertainty $\pm 0.1$ 
in the above exponent.
The somewhat shallower best-fit slope to the observational data ($1.3 \pm 0.3$, right red line)
is likely because the
short-lived FU Ori bursts are not sampled observationally given the small
number of objects.
On the other hand, our numerical models produce FU-Ori-like mass accretion bursts 
(see Fig.~\ref{fig3}), which effectively increase $\mdotavg$ and steepen the model best-fit slope.

\begin{figure}
  \resizebox{\hsize}{!}{\includegraphics{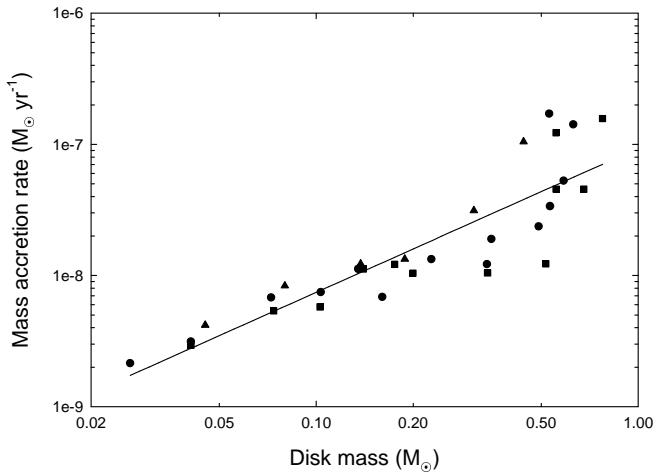}}
      \caption{
Time-averaged mass accretion rate $\mdotavg$ versus
time-averaged disk mass $\mdiskavg$ for all models.
The solid line shows the least-squares fit to the data points.
Symbols have the same meaning as in Fig.~\ref{fig1}.}
         \label{fig2}
\end{figure}

It should be noted that our model accretion rates are derived at 5~AU (the sink cell), 
while the observed accretion rates are measured near the stellar surface.
However, our numerical simulations indicate that accretion rates,
averaged over many orbital periods, vary little with radius in the inner several tens of AU. 
Some physical processes operating in the inner several AU and unaccounted in our numerical 
modeling, will allow accretion onto the stellar surface. The inner
accretion rate is expected to match our calculated value when
time-averaged, even though it may have significant short term variability.

To quantify the characteristic range of accretion rates obtained in our models, 
Figure~\ref{fig1} also shows, for each model, the upper and lower
bounds on the mass accretion rate $\dot{M}$. These values are obtained by smoothing 
the accretion rates over $10^4$~yr periods (to reduce noise) and 
typically correspond
to the accretion rates at the beginning of the T Tauri phase (upper bound) and at
the terminal point of the simulations (lower bound). 
It is evident that our models can account for the observed range of
accretion rates for TTS 
with masses above $1.0~M_\odot$. On the other hand, the observed range 
of accretion rates for  
the intermediate-mass TTS ($0.25~M_\odot < M_\ast <1.0~M_\odot$) is greater
than that implied by the temporal evolution of our models. 
We believe that this can be accommodated ultimately by a broader range
of initial cloud configurations than we have studied here due to numerical
limitations.
However, it is remarkable that our models
do cover the middle portion of the observed $\dot{M}-M_\ast$ phase space.

Figure~\ref{fig2} shows the relation between the time-averaged mass 
accretion rate $\mdotavg$ and the time-averaged disk mass $\mdiskavg$, to which the 
least-squares best fit is
\begin{equation}
\mdotavg  = 10^{-7.0} \; \mdiskavg^{1.1}.
\label{relation3}
\end{equation}
The disk masses are that of matter with surface density above a
value 0.1~g~cm$^{-2}$ that typically characterizes the 
disk-to-envelope transition (VB07). 
In our view, equation~(\ref{relation3}) is the most physically meaningful correlation arising from
our simulations of self-consistent disk formation and evolution due
to global gravitational torques. However, there is also a mild
preference for more massive disks around more massive stars. A plot
of $\xi = \mdiskavg/\mstaravg$ versus $\mstaravg$ has a best-fit
$\xi \propto \mstaravg^{0.3 \pm 0.1}$, with values ranging from
$\sim 5$\% at the low mass end to $\sim 35$\% at the high mass end.
This serves to steepen the 
correlation in Figure~\ref{fig1} so that the best fit is
$\mdotavg \propto \mstaravg^{1.7}$.

\begin{figure}
  \resizebox{\hsize}{!}{\includegraphics{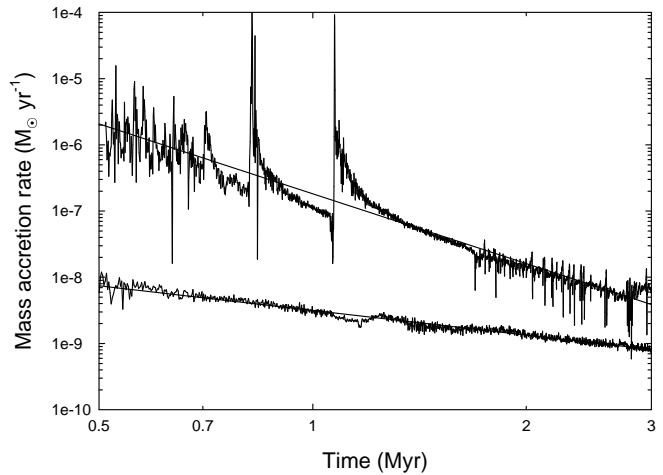}}
      \caption{
Mass accretion rate $\mdot$ versus time for two models, both with $\gamma=\gamma_2$
and initial cloud masses $2.75~M_\odot$ (top) and $0.35~M_\odot$ (bottom).
The solid lines show the least-squares best fits to the data points.}
         \label{fig3}
\end{figure}

Finally, Figure~\ref{fig3} shows the typical mass accretion rates obtained in our models
as a function of time. The top and bottom lines correspond to models with initial cloud masses 
$2.75~M_\odot$ and $0.35~M_\odot$, respectively, both with $\gamma=\gamma_2$. 
The corresponding time-average protostellar masses are $0.63~M_\odot$ and 
$0.026~M_\odot$, respectively. 
The more massive disks clearly 
drive greater mass accretion rates. Furthermore, more massive disks 
show a wider range of mass accretion rates in the T Tauri phase. The least-squares best fit 
to the data in Figure~\ref{fig3} yields an exponent $-3.5$ for 
the $\mdiskavg=0.63~M_\odot$ disk and $-1.2$ for the 
$\mdiskavg=0.026~M_\odot$ disk. 
Massive disks have gravitational torques of greater magnitude, 
which result in greater associated accretion rates. 
The more massive disk is also violently gravitationally unstable in
its early evolution and is characterized by FU-Ori-like accretion bursts 
(VB05; VB06).
We believe that the steeper decline of accretion
rate of the $\mdiskavg=0.63~M_\odot$ disk is caused by the greater effect
of disk self-gravity and possibly the influence of the more massive
stellar object at the center. 

\section{Conclusions}

In this Letter, we have presented models of the late evolution of
self-consistently formed protostellar disks that can explain the
observed correlation of disk accretion rates with stellar mass
for TTS. 
The formation of the disk and its interaction with the surrounding
envelope lead to the development of strong spiral structure 
during the early evolution of the disk (VB05; VB06). 
Even after the former has largely disappeared, low-amplitude 
nonaxisymmetric density
perturbations are sustained in the disk for several Myr.
The gravitational torques due to these
perturbations are sufficient to drive accretion at the rates commonly
inferred around TTS. 
We find that an important property of gravitational
torques is that $\mdotavg$ has an essentially linear dependence on 
$\mdiskavg$. The average disk-to-star mass 
ratio $\xi$ is in the range $\sim 5\% - 35\%$ for models of various
masses, given our adopted range of values of initial cloud angular momentum. 
However, there is a mild trend toward 
greater values of $\xi$ for models with greater masses. The
net result is a best-fit correlation $\mdotavg \propto \mstaravg^{1.7}$,
in good agreement with the observed $\mdot-M_\ast$ relation.

Since the of values of $\xi$ in our models are $\sim 10$ 
times greater than
estimates made from dust emission \citep[e.g.,][]{Andrews,Scholz1}, 
we anticipate two reasons that this discrepancy may be reduced in
the future. Observationally, the inferred gas disk masses may be systematically
underestimated using current techniques \citep[see discussion in][]{Hartmann}.
Theoretically, we need to include additional angular momentum transport
mechanisms such as magnetic braking and the magnetorotational instability.
We note that only a modest increase in the overall average accretion rate in
our current models is required to lead to a dramatic decrease in $\xi$. This is because a small 
relative increase in the stellar mass can significantly reduce the 
much smaller disk mass.

\acknowledgements
The authors thank the referee for providing valuable comments that helped improve the manuscript.
EIV gratefully acknowledges support from an ACEnet Fellowship. 
SB was supported by a grant from NSERC.
We thank Aleks Scholz for helpful discussions, and Samantha Flood and 
Paolo Padoan for providing compilations of observational data from the literature. 
We thank Prof. Martin Houde, the SHARCNET consortium, and the Atlantic Computational 
Excellence Network (ACEnet) for access to computational facilities.


\end{document}